\documentclass[aps,prb,twocolumn,groupedaddress]{revtex4-1}

\usepackage{amsmath}
\usepackage{graphicx}

\usepackage[T1]{fontenc}

\usepackage{hyperref}
\hypersetup{colorlinks=true, citecolor=blue, urlcolor=blue, linkcolor=blue}

\begin{document}

\title{DC conductivity of twisted bilayer graphene: Angle-dependent transport properties and effects of disorder}

\author{M. An\dj{}elkovi\'c}
\email[]{misa.andelkovic@uantwerpen.be}
\author{L. Covaci}
\email[]{lucian.covaci@uantwerpen.be}
\author{F. M. Peeters}
\email[]{francois.peeters@uantwerpen.be}
\homepage[]{http://cmt.uantwerpen.be}

\affiliation{Departement Fysica, Universiteit Antwerpen, Groenenborgerlaan 171, B-2020 Antwerpen, Belgium}

\date{\today}

\begin{abstract}
	The in-plane DC conductivity of twisted bilayer graphene (TBLG) is calculated using an expansion of the real-space Kubo-Bastin conductivity in terms of Chebyshev polynomials. We investigate within a tight-binding (TB) approach the transport properties as a function of rotation angle, applied perpendicular electric field and vacancy disorder.  We find that for high-angle twists, the two layers are effectively decoupled, and the minimum conductivity at the Dirac point corresponds to double the value observed in monolayer graphene. This remains valid even in the presence of vacancies, hinting that chiral symmetry is still preserved. On the contrary, for low twist angles, the conductivity at the Dirac point depends on the twist angle and is not protected in the presence of disorder. Furthermore, for low angles and in the presence of an applied electric field, we find that the chiral boundary states emerging between AB and BA regions contribute to the DC conductivity, despite the appearance of localized states in the AA regions. The results agree qualitatively with recent transport experiments in low-angle twisted bilayer graphene.
\end{abstract}

\pacs{}

\maketitle

\section{\label{section:intro}Introduction}

    The fabrication of multilayer structures is a promising route for developing novel devices engineered for specific purposes, where one has the freedom to choose between metallic, semiconducting or insulating individual layers. Bilayer graphene with its two stacking types, AA and the energetically more stable AB, is one of the most studied examples\cite{mccann_electronic_2013, rozhkov_electronic_2016}. A misalignment between the layers, produces a moir\'e structure, whose electronic properties are greatly affected by the rotation angle. Rotated bilayers can be produced by growing on the C face of a SiC substrate\cite{hass_growth_2008,sprinkle_first_2009,hass_why_2008-1}, by CVD\cite{li_observation_2009, luican_single-layer_2011, yan_angle-dependent_2012}, or by folding single graphene sheets\cite{poncharal_raman_2008, ni_reduction_2008}. Intentional rotation leads to diverse electronic and optical properties, such as a Dirac like spectrum\cite{luican_single-layer_2011}, the renormalization of the Fermi velocity\cite{ni_reduction_2008, luican_single-layer_2011, yin_experimental_2015}, the appearance of low energy van Hove singularities\cite{li_observation_2009, yan_angle-dependent_2012}, and the localization of low energy states\cite{yin_experimental_2015}, which are all angle dependent. Very recently, experimental evidence suggests that strong correlations play an important role in low twist angle bilayers\cite{kim_tunable_2017}. Due to this plethora of interesting properties, the twisting of bilayer graphene can be used as a modulation technique for solid-state devices\cite{koren_coherent_2016, chari_resistivity_2016}. 
    
    A proper theoretical investigation of the transport properties in twisted bilayer, covering consistently all twist angles and disorder effects, is still lacking. The reason is that real-space modeling of the electrical and optical properties of such structures is very demanding. The moir\'e pattern period is related to the twist angle as $L\approx a_0\sqrt{3}/\left(2\sin(\theta/2)\right)$ for $\theta < 30^\circ$ and the angle changes to $\theta^{'} = \pi/3 - \theta < 30^\circ$ for $30^\circ < \theta \leq 60^\circ$. For small rotation angles the period becomes $L\sim 1/\theta$. This results in a large increase of the computational unit cell. For example, if we build a dense Hamiltonian matrix, the requirements of an exact diagonalization for the system of size $L$ scales in memory like $L^2$, and $L^3$ in time. The situation is slightly better for sparse matrices, which is actually the case for TB Hamiltonians, and when we are only interested in a selected part of the spectrum\cite{weise_kernel_2006}. For mesoscopic disordered systems the eigenvalue problem can be solved only for a small number of energies, which relates then to other quantities of interest. There are several different methods available based on  approximate spectral functions such as the Lanczos recursion technique, and the Kernel Polynomial Method (KPM)\cite{weise_kernel_2006}. There are also the recently developed real-space methods for the calculation of the conductivity as given in Refs. [\onlinecite{garcia_real-space_2015}], [\onlinecite{ferreira_critical_2015}], and [\onlinecite{leconte_efficient_2016}]. The great benefit of latter methods is the possibility to include atomistic disorder effects such as charged impurities, modeled by a random on-site potential, or vacancies, and inspect the localization effects at the TB level with an unparalleled accuracy. 
    
    In this paper we will apply the real-space method for the calculation of the DC conductivity\cite{garcia_real-space_2015} in the linear response regime by using the Bastin formula based on the Kubo formalism\cite{bastin_quantum_1971} to TBLG. We will focus on two different types of configurations: small and large rotation angles. Small angles are of interest because of the large moir\'e superlattice, the suppression of the Fermi velocity, the localization of low energy states, and the appearance of conductive topological boundary modes\cite{san-jose_helical_2013} when an electric field is applied perpendicular to the bilayer. Large angles are also peculiar because of the small supercell, and the Dirac like spectrum owing to the layers being effectively decoupled. To the authors knowledge, the only study of vacancy induced effects in twisted bilayer graphene was presented in Ref.~[\onlinecite{ulman_point_2014}] within a DFT formalism and was limited to the high-angle limit. The band-structure and spatial density of states were investigated and it was found that the effect of a vacancy is very similar to the case of single layer graphene.
    
    We find that the DC conductivity at the Dirac point for low-angle TBLG shows an increase as compared to the high-angle (decoupled) TBLG, or AB stacked bilayer graphene. We apply an interlayer potential difference and report a finite value of the conductivity, almost constant over a wide energy range of $\sim 200meV$, around the Dirac point which is due to the presence of boundary conductive modes between the AB and BA regions of the moir\'e pattern. Vacancy induced effects show a quasi-localized behaviour away from the Dirac point, with a shift of the vacancy DOS peak ($\sim 10meV$), depending on the concentration. The conductivity for high-angle TBLG shows that modes at the neutrality point are still delocalized, like in the clean limit. In the low-angle limit, the localization is a consequence of already broken chirality. We also report a significant modification of the low energy spectrum when hopping terms beyond nearest neighbors (both in and out of plane) are included in the TB model. 
    
    The paper is organized as follows. First, the methodology for the calculation of the electronic properties (Sec. \ref{section:method}) is presented, where the method of polynomial expansion of the Bastin formula and the application of absorbing boundary conditions are introduced. Then, we describe the TBLG lattice structure and the TB Hamiltonian in Sec. \ref{section:model}. Finally, simulated results and discussions are presented in Sec. \ref{section:res}, where we applied the methodology to TBLG structures with different twist angles and related the found effects to analogous counterparts in monolayer and bilayer graphene structures.

\section{\label{section:method}Method}

    The methodology developed in Ref.~[\onlinecite{garcia_real-space_2015}] consists of expanding the Bastin formula for the DC conductivity in the linear response regime\cite{bastin_quantum_1971} in terms of Chebyshev polynomials. Starting from the Bastin formula:
    \begin{equation}
        \begin{split}
        	&\sigma_{\alpha\beta}(\mu,T)=\frac{i e^2\hbar}{\Omega}\int_{-\infty}^{\infty}d\varepsilon f(\varepsilon,\mu,T)  \times \\	
        	&\textrm{Tr}\left\langle v_\alpha\delta(\varepsilon-H)v_\beta\frac{d G^+	(\varepsilon)}{d\varepsilon} - v_\alpha\frac{d G^-(\varepsilon)}{d\varepsilon}v_\beta\delta(\varepsilon-H)  
        	\right\rangle,	
            \label{equation:kubo_bastin_formula}
        \end{split}
    \end{equation}
    the $G^\pm$ and $\delta$ functions are expanded. The final expression reads: 
    \begin{equation}
        \begin{split}
             \sigma_{\alpha\beta}(\mu,T)= &\frac{4e^2\hbar}{\pi\Omega}\frac{4\eta^2}{\Delta E^2} \times\\ 
            & \int_{-\eta}^{\eta}d\tilde{\varepsilon}\frac{f(\tilde{\varepsilon},\tilde{\mu},T)}{(1-\tilde{\varepsilon}^2)^2}\sum_{m,n=0}^{M,N}\Gamma_{n m}(\tilde{\varepsilon})\mu_{n m}^{\alpha\beta},
        	\label{equation:kubo_bastin_formula_expanded}
        \end{split}
    \end{equation}
    with:
    \begin{equation}
        \mu_{mn}^{\alpha\beta} = \frac{g_m g_n}{(1+\delta_{n0})(1+\delta_{m0})}Tr\left[v_\alpha T_m(\tilde{H})v_\beta T_n(\tilde{H})\right],
    \end{equation}
    \begin{align}
    \begin{split}
            \Gamma_{mn}(\tilde{\varepsilon}) &= (\tilde{\varepsilon}-in\sqrt{1-\tilde{\varepsilon}^2})e^{in\arccos (\tilde{\varepsilon})}T_m(\tilde{\varepsilon}) \\ 
        	&+(\tilde{\varepsilon}+i m\sqrt{1-\tilde{\varepsilon}^2})e^{-i m\arccos (\tilde{\varepsilon})}T_n(\tilde{\varepsilon}),    
    	\end{split}
    \end{align}
    where $\mu$ is the chemical potential, $T$ the temperature, $\Omega$ area of the structure, $T_{n,m}(\tilde{H})$ Chebyshev polynomial expansions of the scaled Hamiltonian, $f(\tilde{\varepsilon},\tilde{\mu},T)= 1/(1+e^{(\tilde{\varepsilon}-\tilde{\mu})/\tilde{k_B T}})$ is the Fermi-Dirac distribution, and $v_\alpha$ is the velocity component in the $\alpha$ direction given by the Heisenberg expression $\boldsymbol{v}=-i\left[\boldsymbol {l},H\right]/\hbar$, $\boldsymbol {l}$ being the distance vector. The Hamiltonian is scaled such that the spectrum is in the range of $\left[-\eta,\eta\right]$, where $0<\eta\leq1$. In order to mitigate spurious effects due to Gibbs oscillations, the Lanczos kernel, $g_n=\sinh[\lambda(1 - n/N)]/\sinh(\lambda)$\cite{weise_kernel_2006}, is chosen for the calculation, with $\lambda=4$. Spin degeneracy is taken into account. 
    
    The trace of the matrix $v_\alpha T_m(\tilde{H})v_\beta T_n(\tilde{H})$ can be effectively approximated as an average over random phase vectors:\cite{iitaka_random_2004}
    \begin{equation}
        \begin{split}
            Tr\left[v_\alpha T_m(\tilde{H})v_\beta T_n(\tilde{H})\right]&\approx \\ \frac{1}{N_R}\sum_{r=1}^{N_R} & \langle r \left| v_\alpha T_m(\tilde{H})v_\beta T_n(\tilde{H}) \right| r \rangle,
        \end{split}
        \label{equation:random_trace}
    \end{equation}
    with $\left|r\right\rangle=\left|e^{i\phi_i}\right\rangle$ and $\phi_i$ being the random phase in the range $\left[0,2\pi\right)$. The number of random vectors, $N_R$, needed to converge the sum is usually much smaller than the size of the system. For example, in the calculation of conductivity given in Eq.~(\ref{equation:kubo_bastin_formula_expanded}) we typically $N_R=56$ random vectors and around $N=10000$ moments, which for the width of the spectrum of the Hamiltonian, and the kernel used, corresponds to having a resolution of $4meV$.
    
    If we examine the most demanding part, the calculation of the Chebyshev moments -- $\mu_{nm}$, which requires a double polynomial expansion, there are two ways to solve it. One is by expanding $\left|\tilde{r}_{n}\right\rangle = v_\beta T_n\left|r\right\rangle$ and then for each of the vectors $\left|\tilde{r}_{n}\right\rangle $ using the full expansion, $\left|\tilde{r}_{nm}\right\rangle = T_m \left|\tilde{r}_{n}\right\rangle$. In this way we only have to store two vectors that are used in the recursion. The other way is to split the double iteration into two independent parts, $\left|\tilde{r}_{n}\right\rangle = v_\beta T_n\left|r\right\rangle$ and $\left|\tilde{r}_{m}\right\rangle =  T_m v_\alpha\left|r\right\rangle$. This requires the use of a large amount of memory because all the vectors in the iterations must be stored. The first approach can be efficiently used on GPUs, where one can perform fast sparse matrix - vector multiplication. When a large amount of fast storing memory needs to be used, this can only be done on the RAM, therefore the second approach is more suitable for CPUs. A mixed approach can use benefits of both methods, where we can save as many vectors as we have available memory on the GPU. 
    
    At zero temperature the set of equations simplifies, and the calculation of the longitudinal conductivity can be done with the same $O(N^\star)$ complexity\cite{leconte_efficient_2016} ($N^\star=N \times N_A \times N_R$, $N$ being the number of moments, $N_A$ the number of atoms and $N_R$ the number of random vectors) as needed for the DOS expansion\cite{weise_kernel_2006}. We will use this approach in the following to calculate the longitudinal DC conductivity. 
	
    \subsection{\label{section:abs}Absorbing boundary conditions}

        When we want to study only bulk properties of materials, a very large system or appropriate boundary conditions are needed. As the maximal size of the system depends on the computational resources available, it is usually suitable to impose boundary conditions rather than to increase the system size. The periodicity of the structure is relatively easy to implement for basic structures in the TB model, as we need only a well defined unit cell. The problem appears for more complex structures, as is the case of TBLG with an arbitrary twist angle, for which the unit cell in general is not well defined in particular for the case of incommensurability of the two layers. When a magnetic field is added, it breaks the lattice translational symmetry. In this case the periodicity of the magnetic vector potential and the unit cell of the TBLG must match, which imposes on top of discrete values of the field a specific periodic gauge for the vector potential. However, if we apply the so-called absorbing boundary conditions the above limitations are lifted, and the calculation can be done for arbitrary magnetic field. A drawback is in the slight increase of the calculation time because of two additional matrix-vector multiplications, as shown below. 
        
        One may say that none of these choices is perfect, because periodic boundary conditions give discrete states around the Dirac point and increase the number of the sampling $\boldsymbol{k}$ points in the inverse space needed, while absorbing boundary conditions can give false reflections, which results in a finite increase of low energy states. Depending on the particular purpose, one should choose the most appropriate approximation, which in the case of twisted bilayer graphene is the latter one.
        
        The procedure for the application of absorbing boundary conditions is the following \cite{munoz_disordered_2015}. The recursion relation for the regular Chebyshev expansion reads: 
        \begin{equation}
        	\left|r_n\right\rangle = 2\tilde{H}\left|r_{n-1}\right\rangle - \left|r_{n-2}\right\rangle,
        	\label{equation:chebyshev_req}
        \end{equation}
        which in the presence of the absorbing boundary is changed to\cite{grozdanov_recursion_1995}:
        \begin{equation}
        	\left|r_n\right\rangle = e^{-\gamma}\left[2\tilde{H}\left|r_{n-1}\right\rangle - e^{-\gamma}\left|
        	r_{n-2}\right\rangle\right]
        	\label{equation:chebyshev_req_abs}
        \end{equation}
        in order to account for the damping of the wave functions at the edges of the system.
       
        The initial conditions for the modified recursion relation in Eq.~(\ref{equation:chebyshev_req_abs}) are now:
        $\left|r_{0}\right\rangle = \left|e^{i\phi_i}\right\rangle$ and $\left|r_{1}\right\rangle = e^{-\gamma}H\left|r_{0}\right\rangle$. We also set the initial random vectors in Eq.~(\ref{equation:random_trace}) to have non-zero elements only in the region where there is no absorption, and therefore the reconstruction of the expanded functions \cite{grozdanov_recursion_1995,munoz_disordered_2015} is analogous to the reconstruction obtained in the absence of the absorbing potential.
        
        The function $\gamma(x)$ is a position dependent damping factor adopted from Refs.~[\onlinecite{manolopoulos_derivation_2002}] and [\onlinecite{gonzalez-lezana_quantum_2004}] as $\gamma(x) = \sinh^{-1}[V(x)]$. $\Gamma(x)=-iV(x)$ is defined as the complex absorbing potential (CAP): 
        \begin{equation}
        	\Gamma(x)=-i\left(\frac{2\pi}{\Delta x}\right)^2 f(x^\prime),
        	\label{equation:absorbing_potential}
        \end{equation}
        where $\Delta x = x_2 - x_1$ is the width of the absorbing region. Finally, $f(x^\prime)$ is non-zero only between $x_1$ and $x_2$:
        \begin{equation}
        	f(x^\prime)=\frac{4}{(c-x^\prime)^2} + \frac{4}{(c+x^\prime)^2} - \frac{8}{c^2},
        	\label{equation:absorbing_function}
        \end{equation}
        where $x^\prime=c(x-x_1)/\Delta x$, while $c=2.62$ is a numerical constant. Similar absorbing potential can be defined in a cylindrical geometry, where $\gamma$ depends only on the radial direction and is non-zero in the absorbing region.
        
        Two different vectors in Eq.~(\ref{equation:chebyshev_req_abs}) need to be multiplied by the absorbing function, but we can simplify and remove one product with the following modification of the Hamiltonian $\tilde{H}_{abs} = 2e^{-\gamma}H$: 
        \begin{equation}
        	\left|r_n\right\rangle = \tilde{H}_{abs}\left|r_{n-1}\right\rangle - e^{-2\gamma}\left| r_{n-2}\right\rangle.
        	\label{equation:chebyshev_req_modified}
        \end{equation}	

\section{\label{section:model}Model}

    Following Refs. [\onlinecite{Trambl-2010}] and [\onlinecite{van_der_donck_piezoelectricity_2016}] we briefly discuss the geometrical properties of twisted bilayer graphene. The structure consists of two graphene layers, and we start with AA stacked bilayer graphene, which is rotated over an arbitrary angle $\theta$ with respect to an atomic site. A rotation of $\pi/3$ results in AB stacked bilayer graphene. After the rotation, the lattice vectors of the different layers become: 
    \begin{equation}
        \begin{split}
            \boldsymbol{a}_{1,2}&=a_{0}R(\theta/2)(3/2,\pm \sqrt{3}/2, 0), \\ 
            \boldsymbol{a}_{1,2}'&=a_{0}R(-\theta/2)(3/2,\pm \sqrt{3}/2, 0)
        \end{split}
    \end{equation}
    where $a_0=0.142nm$ is the carbon-carbon distance, $R(\theta)$ the rotation matrix and $\boldsymbol{a}_{1,2}$ and $\boldsymbol{a}_{1,2}'$ are lattice vectors of bottom and top layer, respectively. The position of the carbon atoms ($A/B$ in the bottom and $A'/B'$ in the top layer) are defined as:
    \begin{equation}
        \begin{split}
            \boldsymbol{r_n}^A&=n\boldsymbol{a_1}+m\boldsymbol{a_2}, \\
            \boldsymbol{r_n}^B&=n\boldsymbol{a_1}+m\boldsymbol{a_2} + \boldsymbol{\delta}, \\ 
            \boldsymbol{r_n}^{A'}&=n\boldsymbol{a_1'} + m\boldsymbol{a_2'} + \boldsymbol{\delta_z}, \\
            \boldsymbol{r_n}^{B'}&=n\boldsymbol{a_1'} + m\boldsymbol{a_2'} + \boldsymbol{\delta'} + \boldsymbol{\delta_z}, \\ 
        \end{split}
    \end{equation}
    and $\boldsymbol{\delta}=a_0R(\theta/2)(1,0,0)$, $\boldsymbol{\delta'}=a_0R(-\theta/2)(1,0,0)$, $\boldsymbol{\delta_z} = d_0(0,0,1)$ where $d_0=0.335nm$ is the interlayer distance.
    
    We simulated a circular sample with a radius of $R=80nm$. The width of the absorbing boundary region is taken to be $5\%$ of the radius, with a small additional region of complete absorption, i.e. $e^{-\gamma}=0$, around the edges of the sample.
    
    We start from a TB Hamiltonian for a non-interacting system:    
    \begin{equation}
    	H = -\sum_{i,j}t(\boldsymbol{r}_i,\boldsymbol{r}_j)c_i^{\dagger}c_j + 
    	\sum_i \varepsilon_i c_i^{\dagger}c_i + 
    	\sum_i \Delta_i c_i^{\dagger}c_i,
    \end{equation}
    where $\varepsilon_i$ is used to model the onsite disorder, and is chosen from an uniform random distribution on the interval $[-0.15eV, 0.15eV]$, while $\Delta_i$ is the applied on-site potential. The transfer integral between the sites is taken as in Refs. [\onlinecite{trambly_de_laissardiere_numerical_2012}] and [\onlinecite{moon_energy_2012}]:
    \begin{equation}
        \begin{split}
            -t(\boldsymbol{r}_i,\boldsymbol{r}_j) =V_{pp\pi}\left[1-\left(\frac{\boldsymbol{d\cdot e}_z}{d}\right)^2
        	    \right] + V_{pp\sigma}\left(\frac{\boldsymbol{d\cdot e}_z}{d}\right)^2, 
	    \end{split}
	\end{equation}
	\begin{equation}
        \begin{split}
            V_{pp\pi} &= V_{pp\pi}^0e^{-\frac{d-a_0}{\delta}}F_c(d), \\
            V_{pp\sigma} &= V_{pp\sigma}^0e^{-\frac{d-d_0}{\delta}}F_c(d),
    	    \label{equation:hoppings}
        \end{split}
    \end{equation}
    with $V_{pp\pi}^0=-2.7eV$ and $V_{pp\sigma}^0=0.48eV$ are intralayer and interlayer hopping integrals, respectively. $\boldsymbol{d}$ is the vector between two sites, $d$ is the distance between them, and $\delta = 0.3187 a_0$ is chosen in order to fit the next-nearest intralayer hopping to $0.1V_{pp\pi}^0$. By comparing the low energy spectrum we conclude that the interlayer hoppings between sites further away than the next-nearest neighbor distance, and further from the distance of $1.5d_0$ in the interlayer case, contribute only to small energy shifts of the spectra, without modifying it significantly. Due to this fact, we applied a Heaviside $\theta_c$ cutoff function:
    \begin{equation}
        F_c(d) = 
        \begin{cases}
            \theta_c(d-1.5d_0),& dz \geq 0.5d_0\\
            \theta_c(d-1.1\sqrt{3}a_{0}),  & dz < 0.5d_0.
        \end{cases}
        \label{equation:cutoff_function}
    \end{equation}
    Neglecting the above mentioned hoppings resulted in a considerable gain in calculation time due to a reduced number of non-zero elements in the Hamiltonian. 
    
    Long-range interlayer coupling breaks the electron/hole symmetry in the TB Hamiltonians and results in a shift of the Dirac point energy ($E_D$). In calculations where the interlayer neighbors beyond the nearest ones are included, the energy axis is shifted with $E_{D0}=3t_{nn}=0.3V_{pp\pi}^0=0.81eV$, i.e. the energy of the Dirac point in monolayer graphene when next-nearest neighbors are included.
    
    \subsection{\label{subsection:TBLG}Twisted bilayer graphene}
        To visualize the stacking order in the TBLG structure we use a modification of the registry index (RI)\cite{marom_stacking_2010, hod_registry_2013}, which is a geometrical parameter for quantifying the matching between different layers.         The RI is normalized to a range between the RI of AB stacked bilayer, $RI=0$, and AA stacked graphene, $RI=1$.
        
        One important aspect to consider is the possibility of lowering the energy of the bilayer due to the van der Waals interaction between the layers. We, therefore, in order to find the relaxed bilayer structure, performed an energy minimization through the use of molecular dynamics simulations. For in-plane interactions we use the Brenner potentials, while for the out-of-plane van der Waals interaction we use the registry dependent Kolmogorov-Crespi potential\cite{Crespi_int_pot_graphittic} with parameters given in Ref.~[\onlinecite{Yazyev_relax}]. The relaxation results in a configuration that lowers the van der Waals energy through the maximization of the AB regions, which have the lowest energy, at the expense of in-plane strain\cite{wijk_relaxation_2015}.
        
\begin{figure}
        	\includegraphics[width=\columnwidth]{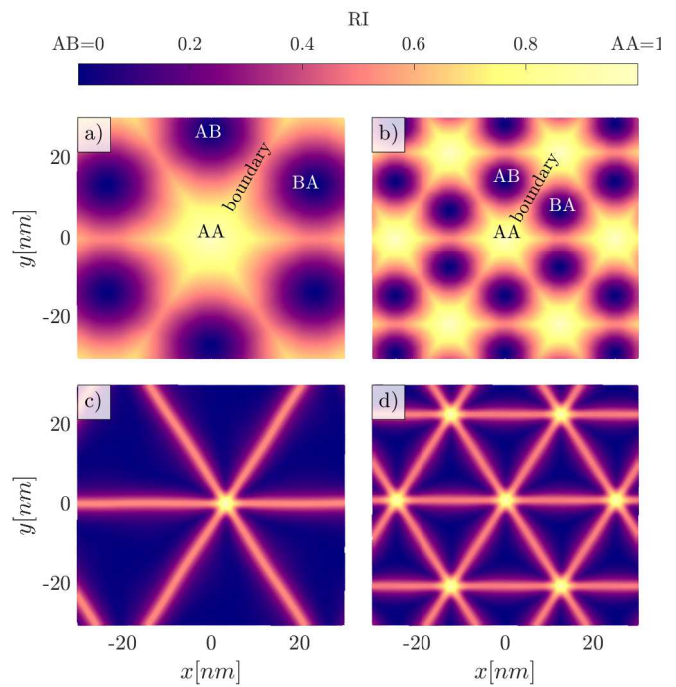}
        	\caption{\label{figure:TBLG_RI} Spatially dependent registry index for TBLG with twist angle: a) $\theta=0.3^\circ$, b) $\theta=0.565^\circ$, and after relaxation for $\theta=0.3^\circ$ and $\theta=0.565^\circ$, in c) and d) respectively.}
        \end{figure}

        After calculating the registry index for each atom, we plot in Fig.~\ref{figure:TBLG_RI} the minimal value between the A and B sublattices. In Figs. \ref{figure:TBLG_RI}(a) and \ref{figure:TBLG_RI}(b) we see wide regions of nearly AA stacked graphene, which become suppressed after the structure is relaxed, in favor of the AB and BA regions, see Figs. \ref{figure:TBLG_RI}(c) and \ref{figure:TBLG_RI}(d). 	
    
	    Different regimes depending on the twisting angle exist. Rotations by angle in the range $15^\circ - 30^\circ$ (high angles) result in TBLG behaving as two decoupled layers, both with linear dispersion near the Dirac point\cite{luican_single-layer_2011}. Lower angles in the range $2^\circ - 15^\circ$ (intermediate angles) still show a  linear dispersion but the velocity is renormalized, i.e. it is suppressed. For very small rotation angles in the range of $0^\circ < \theta < 2^\circ$ (low angles) the low energy spectrum is no longer linear and the velocity has an oscillating behaviour with a considerable suppression at specific angles of rotation, $1.13/n$ with $n$ an integer, obtained from a TB Hamiltonian in Ref. [\onlinecite{trambly_de_laissardiere_numerical_2012}], resulting in the localization of the electronic states in the AA regions \cite{trambly_de_laissardiere_numerical_2012, Trambl-2010, bistritzer_moire_2011-1}. The resulting angles are slightly different from those introduced in Ref.~[\onlinecite{bistritzer_moire_2011-1}], where an effective continuum model was used. TBLG with $30^\circ < \theta < 60^\circ$ results in the same unitcell as the one in the range of $0^\circ < \theta < 30^\circ$. We first investigate the density and transport properties of TBLG for a purpose of making an unified description of different regimes with an arbitrary twist angle.
	    
    	In structures with rotation angles below $1^\circ$, AB, BA and AA-like bilayer regions are easily distinguishable because of the large size of the moir\'e supercell. When a perpendicular voltage is applied, a gap opens in the AB and BA regions\cite{zhang_direct_2009, mccann_electronic_2013, rozhkov_electronic_2016}, while the localization of the states in the AA region is enchanced\cite{yin_experimental_2015}, where the voltage effectively changes the intralayer hopping\cite{rozhkov_electronic_2016}. The topological change between the AB and BA regions induces the appearance of topological boundary states, which were theoretically predicted\cite{martin_topological_2008,san-jose_helical_2013, zhang_valley_2013, vaezi_topological_2013}, and only recently experimentally confirmed\cite{ju_topological_2015,yin_direct_2016} in the AB bilayer graphene domain walls. For these states to appear in TBLG, large regions of nearly AB stacked bilayer graphene and/or high voltage are required\cite{san-jose_helical_2013}. 
        
        In the following we examine the DC conductivity near the Dirac point and the effect of vacancies, as strong short-range scatterers, as well as the appearance of boundary states and their influence on the conductivity of TBLG structures with small rotation angles. 
        
\section{\label{section:res}Results and discussion}
    
    \subsection{Angle dependent conductivity of plain TBLG}

\begin{figure}
        	\includegraphics[width=\columnwidth]{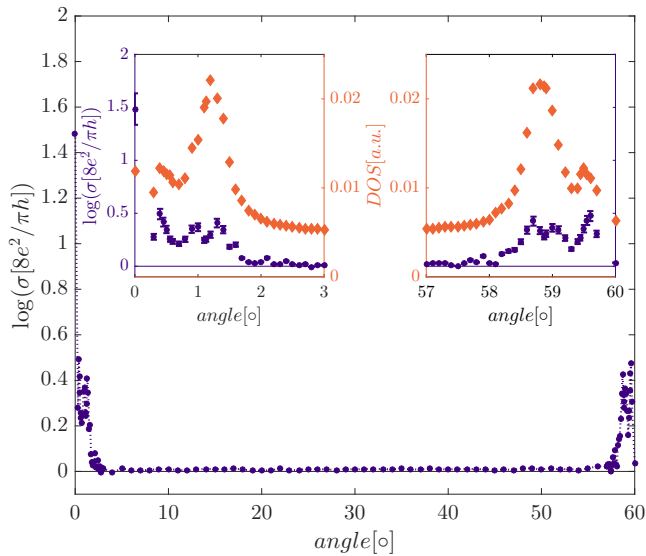}
        	\caption{\label{figure:Cond_rotation} Conductivity at the Dirac point as a function of the rotation angle. The conductivity of decoupled layers is $2\sigma_0 =  8e^2/\pi h$. The inset shows the conductivity and DOS of structures with $\theta\in[0^\circ,3^\circ]$ (left inset) and $\theta\in[57^\circ,60^\circ]$ (right inset). The shape of the conductivity as a function of the rotation angle follows qualitatively the change in DOS (orange diamond markers) at the neutrality point. Error bars shown together with the conductivity values in the insets account for the estimate of the maximal error in the calculation (<$10\%$).}
        \end{figure}

    	As a reference point, we calculated the conductivity of monolayer graphene and high-angle TBLG with $\theta=21.78^\circ$, which resulted, respectively, in the known value of $\sigma_0 = 4e^2/\pi h$ and $8e^2/\pi h$, at the neutrality point. This result proves that the high-angle TBLG at low energies behaves as two decoupled monolayers. Next, we explore the conductivity of TLBG as a function of twist angle, from low to high. Fig.~\ref{figure:Cond_rotation} shows the DC conductivity of TBLG at the Dirac point for different rotation angles, from AA bilayer graphene ($\theta=0^\circ$) to AB ($\theta=60^\circ$). For the angles in the range $\theta\in[2^\circ,58^\circ]$ the constant value of $2\sigma_0$ means that layers are decoupled for a broad range of angles. In this regime, the minimal value of the conductivity is given by the pseudo-diffusive transport that is a consequence of the chirality of the carriers and is protected by the chiral symmetry.
        
        A very different regime emerges at low-angles where we find that the conductivity at low energies is strongly enhanced as compared to the decoupled case. In order to clarify this further, we plot in Fig.~\ref{figure:Dos_angle_with_and_without_V}(a)($V=0V$) the DOS near the Dirac point as a function of rotation angle. For high-angle of rotation, the DOS of decoupled layers is found to be linear as a function of energy, together with a shift of the van Hove peaks towards the Dirac point with decreasing twist angle. For angles below $\sim 1.5^\circ$, states with a small Fermi velocity which are restricted to the AA stacked regions appear at the Dirac point due to the merging of van Hove singularities\cite{Trambl-2010, Santos_continuum}. We might call these states even quasi-localized, as their velocity is not completely vanishing.
        
        We explain the increase in conductivity at low-angles through the finite density of states in regions other than the AA (opposed to the vanishing density in the decoupled case), and a finite non-zero, albeit much smaller compared to the monolayer, value of the Fermi velocity, together with the large density of the AA states that contribute to the conductivity. This fact will be proven in the latter section, where we consider the effect of the applied interlayer potential which will further localize the AA states. There is an oscillating behaviour (shown in the insets of Fig.~\ref{figure:Cond_rotation}), with a slight suppression of the conductivity at the angles that are close to the so-called magic angles (as detailed in Ref.~[\onlinecite{trambly_de_laissardiere_numerical_2012}]). Our calculation suggests that the enhancement in the density of states prevails over the suppression of the velocity. As we can see, the value of conductivity follows the shape of DOS at the neutrality point. 
        
        The conductivity of the two limiting cases, i.e. the AB and AA alignments, is worth mentioning. We obtain values of $\sim 2\sigma_0$ (AB stacked) and high value of $\sim 60\sigma_0$ (AA stacked). Our findings agree with an increase in the DC limit of the optical conductivity of AA stacked bilayer as found in Refs. [\onlinecite{tabert_dynamical_2012}] and [\onlinecite{rozhkov_electronic_2016}] where a strong Drude peak appearing at low frequencies was reported. The reasons behind the value of the Dirac conductivity of AB stacked graphene are twofold. The first one is the Lifshitz transition due to skew interlayer hopping that induces the trigonal warping and increases the value of the conductivity to $6\sigma_0$\cite{cserti_trigonal, koshino_transport_2006, mccann_electronic_2013}, which we implicitly take into account in Eq.~(\ref{equation:hoppings}). The second reason that changes the minimal conductivity is the disorder\cite{koshino_transport_2006, mccann_electronic_2013}, which broadens the low energy states caused by the Lifshitz transition, and hence decreases the conductivity. In addition, absorbing boundary conditions effectively act as disorder contributing to the broadening of the states, finally resulting in the decrease of the conductivity.
            
        In the following we further apply the methodology to specific structures, i.e. the low-angle TBLG in applied electric field, and low and high-angle TBLG with vacancy disorder. 
	
    \subsection{Low-angle TBLG in perpendicular electric field}
 
    	An electric potential difference applied to the TBLG layers depletes the low energy states in the AB regions due to the appearance of an induced electronic gap whose size is experimentally tunable. In addition, topological changes between AB and BA regions are expected to give rise to gapless boundary states protected from valley mixing, which can act as transport channels\cite{vaezi_topological_2013}, and should preserve the non-zero conductivity.
    	
\begin{figure}
    	    \includegraphics[width=\columnwidth]{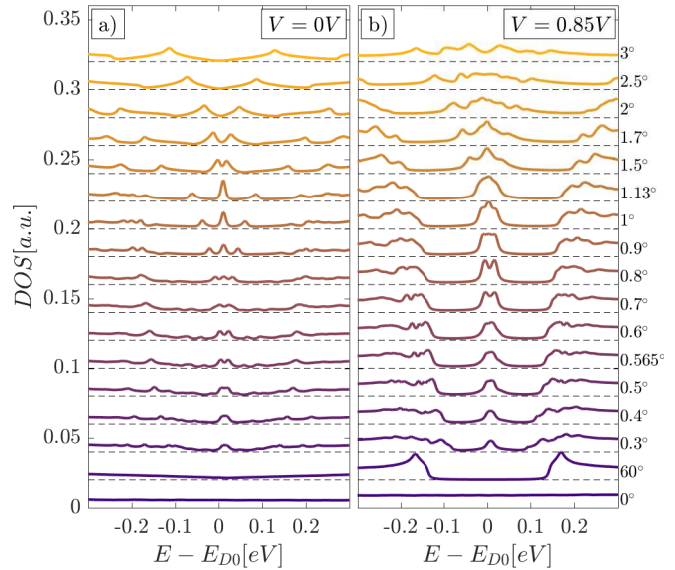}
    	    \caption{\label{figure:Dos_angle_with_and_without_V} Low energy DOS for different rotation angles a) without, and b) with applied interlayer potential. The Dirac point exhibits a shift on the order of $\sim meV$ when comparing two different angles, originating from a slight change in interlayer hoppings. The merging of the van Hove singularities can be seen at lower energies with decreasing angle. Below $1.5^\circ$ the increased density (and decreased velocity) of states at the Dirac point is a result of quasi-localized states in the AA region.}
        \end{figure}

    	First, we present the density of states for small-angle TBLG without and with an applied interlayer potential difference $V=0.85$V in Fig.~\ref{figure:Dos_angle_with_and_without_V}(a)  and Fig.~\ref{figure:Dos_angle_with_and_without_V}(b), respectively. The applied voltage ($V=0.85V$) is chosen to result in the maximal size of the gap of bulk AB stacked graphene\cite{rozhkov_electronic_2016}. By comparing different structures, we see on one hand that by decreasing the angle, a merging of the low energy van Hove singularities occurs and results in an increased density around the Dirac point (AA states). On the other hand, by applying a voltage difference (the DOS for AB stacked graphene shows a large gap on the order of $0.3eV$), the states are depleted in the AB like regions. 
    	
\begin{figure*}
        	\includegraphics[width=\columnwidth]{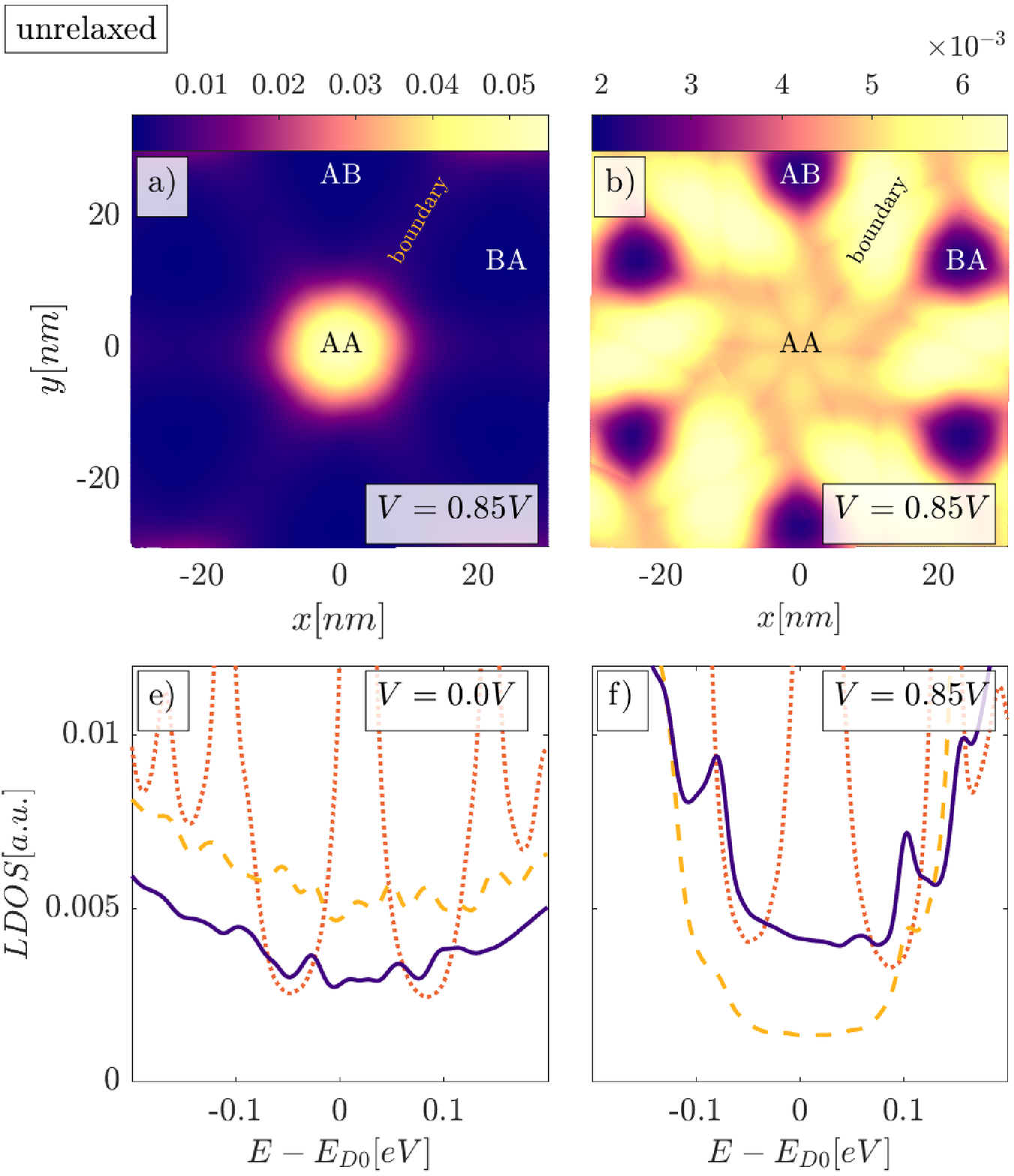}
        	\includegraphics[width=\columnwidth]{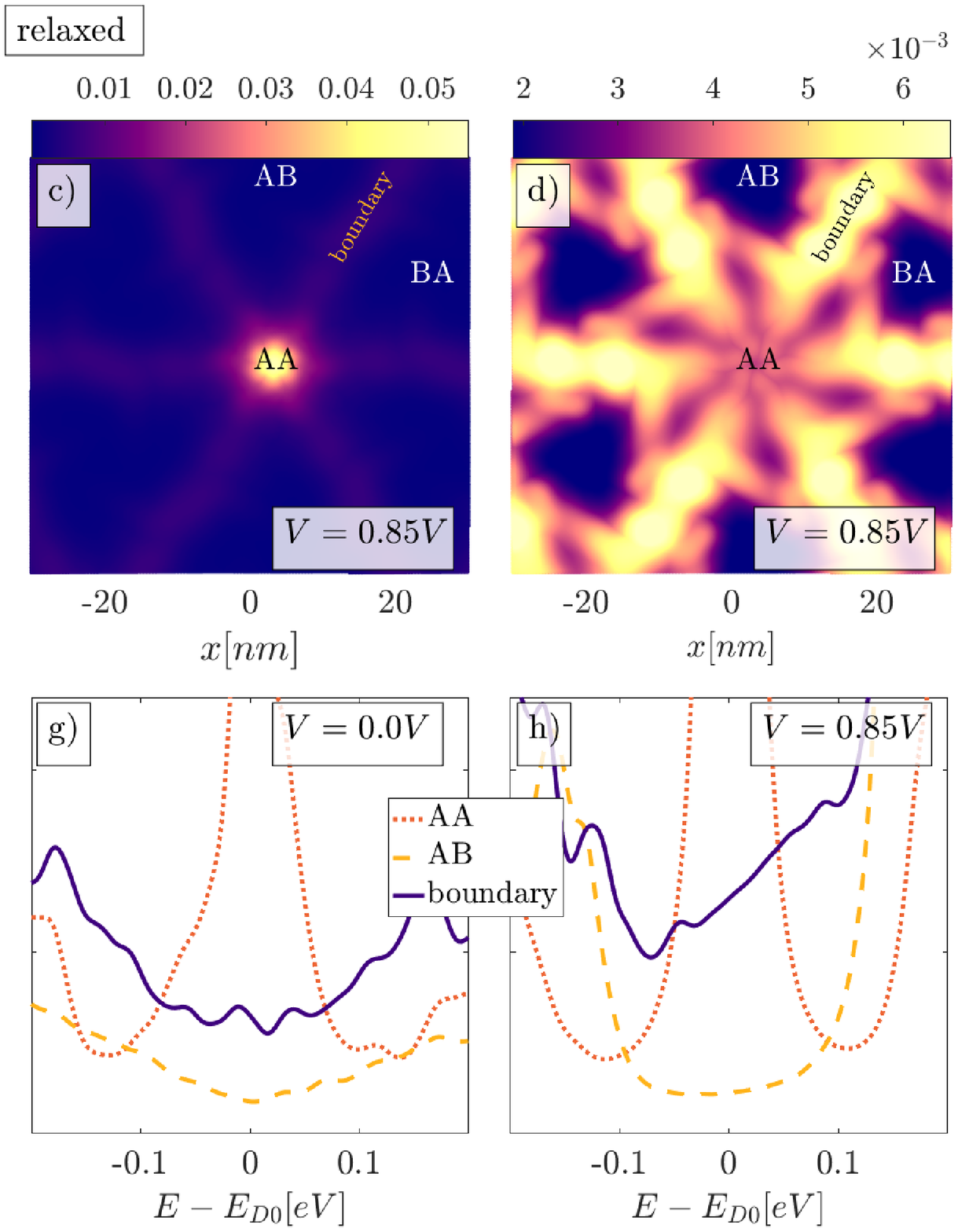}
        	\caption{\label{figure:Ldos_map_both_0_3} Local density of states for $\theta=0.3^\circ$ angle TBLG before a), b), e), f) and after relaxation c), d), g), h). Spatially dependent LDOS maps with the applied interlayer potential $V=0.85V$, are shown at energies a) $E=0.01eV$, b) $E=-0.06eV$, c) $E=0.01eV$, and d) $E=-0.06eV$. LDOS at different central sites in AB, AA, and AB/BA boundary region are presented in order to differentiate the dominant states in the low energy spectrum. The LDOS is plotted at sites belonging the dimer pair. Insets e), g) show LDOS without ($V=0V$) and f), h) with ($V=0.85V$) the applied potential. Movies for the LDOS maps for a wide range of energies are provided as supplemental materials.}
        \end{figure*}

    	The finite DOS at the energies where AB regions are gaped, and away from the localized AA peak, is predominantly the result of the boundary states going from an AA region to the neighboring one. Both the localization in the AA regions and the boundary states can be clearly seen in the spatial LDOS map of generic and relaxed $\theta=0.3^\circ$ structures for $V=0.85V$ at different energies, see Fig.~\ref{figure:Ldos_map_both_0_3}. Because of the finite area of the AB regions the local density of states in those regions is finite but strongly suppressed when compared to the density of the boundary states.
    	
        \begin{figure}[t]
        	\includegraphics[width=\columnwidth]{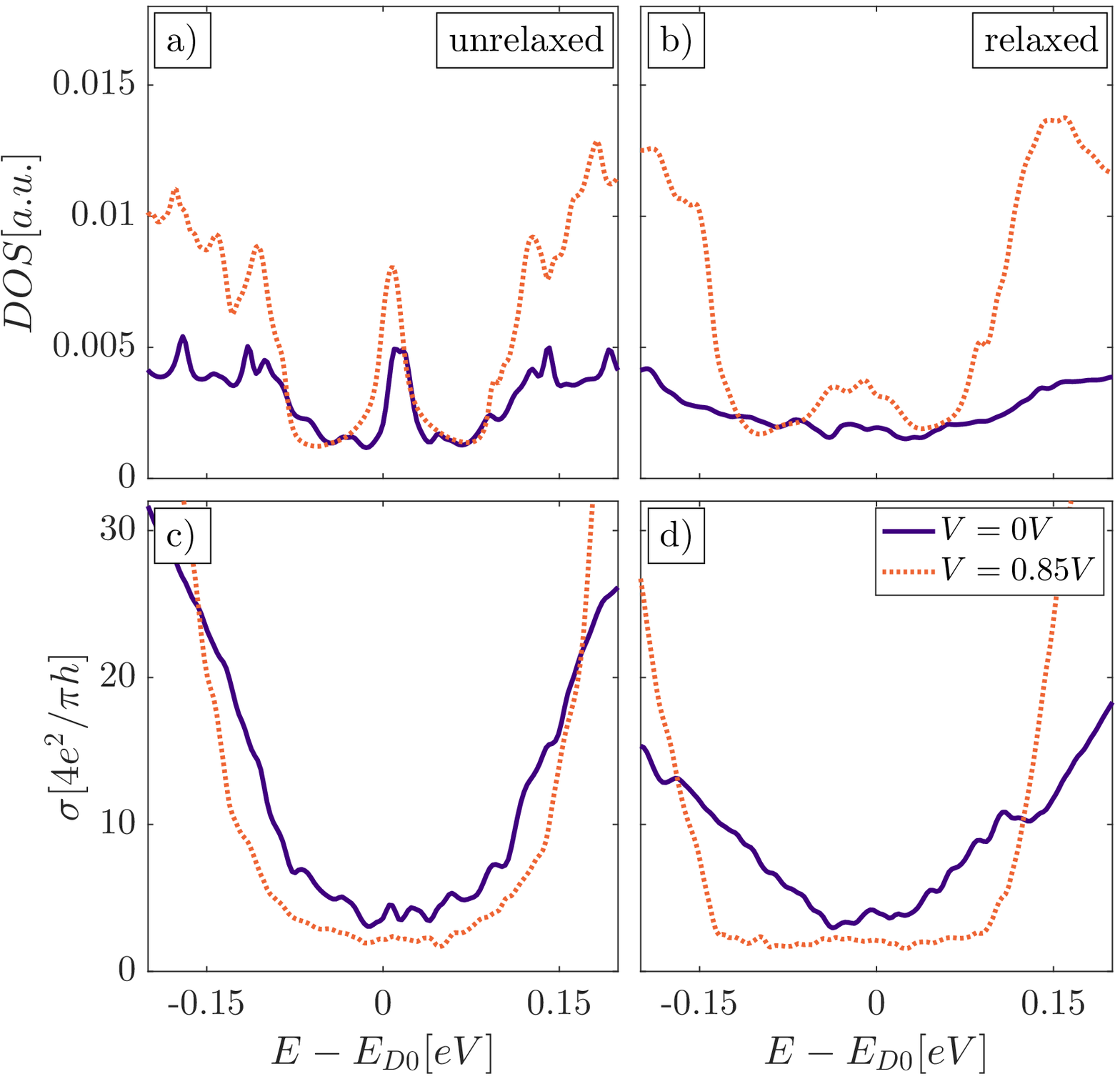}
        	\caption{\label{figure:Conductivity_low_angle_TBLG_0_3} Spectral properties of  $\theta=0.3^\circ$ TBLG without a) and c), and with relaxation b) and d). Panels a) and b) are showing DOS, while panels c) and d) show the low energy DC conductivity in the presence and absence of the applied interlayer potential.}
        \end{figure}
        In Fig.~\ref{figure:Conductivity_low_angle_TBLG_0_3} we compare the DOS and DC conductivities of generic and relaxed structures for TBLG with $\theta=0.3^\circ$ in the presence and absence of a perpendicular electric field. The non-zero conductivity when an electric field is applied, showed in Figs.~\ref{figure:Conductivity_low_angle_TBLG_0_3}(c, d), is a consequence of the boundary region states (between AB and BA regions) whose both density and velocity are dominant in the low energy range. To further prove this claim, let us consider in more details how are different spatial regions affected by the electric field (Fig.~\ref{figure:Ldos_map_both_0_3}(e-h)). 
        As stated before, the depletion of the states in the AB like regions is a sign of the gap opening, which only in the case of perfect AB bilayer would be total. In TBLG, we have the finite (but negligible compared to other regions) density influenced by the neighboring regions. An additional effect is that the AA states are more isolated, concluded from the fact that although their density is increased (Fig.~\ref{figure:Dos_angle_with_and_without_V}), we see a drop of the conductivity at the Dirac point, meaning that effectively a stronger localization is induced. AB/BA boundary regions host now the topological states which are the cause of LDOS increase at the boundary sites, see Fig.~\ref{figure:Ldos_map_both_0_3}(f, h) compared to Fig.~\ref{figure:Ldos_map_both_0_3}(e, g). We observe a nearly flat value of the conductivity in the energy range corresponding to the gap appearing in the AB regions. The correlation between the LDOS and the conductivity is not apparent before making the distinction between localized modes (resonances in the LDOS) and a continuum background\cite{san-jose_helical_2013} responsible for transport. On average (DOS shown in Fig.~\ref{figure:Conductivity_low_angle_TBLG_0_3}(a, b)), this background, whose monotonousness is influenced by the sublattice and layer polarization of the LDOS as well, result in the almost constant value of conductivity. 
        
        In order to further promote the appearance of boundary states, and enhance the low energy conductivity, we should enlarge the energetically favourable AB (gaped) regions, while on the other hand decrease the localized AA regions. This can be achieved by relaxing the low-angle structures, thus by minimizing the van der Waals interaction. The results shown in Fig.~\ref{figure:Ldos_map_both_0_3} confirm our goal, we see well defined boundary states and narrowing of the AA regions. The flatness of the conductivity is now on a broader energy range, which is a result of the gap increase in the enlarged AB regions (Fig.~\ref{figure:Conductivity_low_angle_TBLG_0_3}(d)). The decrease of the DOS at the Dirac point did not result in a significant decrease of the conductivity, further confirming the statement on the increased localization, and hence small contribution of the AA states in the presence of the applied electric field. 
    	
        Recent experiments on low-angle TBLG show that electron-electron interactions have a strong effect on the transport properties in flakes with accurately controlled angles\cite{kim_tunable_2017}. The interactions are reported to affect the transport properties only when the width of the flat bands localized in the AA region is comparable to the energy scale of the interactions. Although in our work interactions are not considered, the validity of our results stand when interactions are not significant. Our results agree qualitatively with the experimental findings at finite temperature and for specific angles, where the interactions induced gaps are closed. At rotation angle close to the second magic angle ($\theta \sim 0.5^\circ$) the correlations are not altering the transport properties (Fig.~3(b) for ($\theta \sim 0.43^\circ$) in Ref.~[\onlinecite{kim_tunable_2017}]), which is also the case for the higher temperature conductance measurements (Fig.~1(f) $T=80K$ in Ref.~[\onlinecite{kim_tunable_2017}]). These measurements agree with the description of nearly flat conductivity at low energies obtained from our calculation. Our claim is that the main contribution to the conductivity comes from the boundary region states, which results in an almost constant DC conductivity.
        
    \subsection{Disorder effects}
        
        Long-range impurities in graphene can result in a localization only when the induced disorder potential is strong enough to affect the intervalley scattering rates\cite{zhang_localization_graphene_2009}. More interesting are the topological defects that are the limiting factor for the transport even for single realization\cite{zhang_quantum_blockade_loop_2008}. Vacancies, as short-range resonant scatterers, form quasi-localized states in the vicinity of missing atoms, appearing in the energy spectrum near the Dirac point, with their amplitude decaying as $\sim 1/r$\cite{pereira_modeling_2008}. They can result in the scattering of carriers over large momentum and induce intervalley scattering, which, if strong enough, would turn graphene into an Anderson insulator\cite{mucciolo_disorder_2010}. A remarkable feature of vacancy induced zero energy mode (ZEM) in monolayer graphene is the protection provided by the chiral symmetry at the Dirac point manifested as the preserved universal value of conductivity over a wide range of disorder concentrations\cite{peres_colloquium_2010, ferreira_critical_2015}, which is the known value for the conductivity of clean graphene in the thermodynamic limit\cite{katsnelson_zitterbewegung_2006}, and shows that vacancy induced ZEM is delocalized. On the other hand, the vacancy states are in particular sensitive to electron-hole symmetry. In the case of monolayer graphene, the asymmetry induces a shift of the vacancy peak away from the Dirac point to the continuum of the extended states\cite{pereira_disorder_2006,pereira_modeling_2008}. 
        
        In the following we investigate the effects of vacancies on the low-angle $\theta=1.13^\circ$ and high-angle $\theta=21.78^\circ$ TBLG conductivity. All the values obtained are an average of $6$ randomly distributed vacancy realizations. 
          
\begin{figure}[htb]
    	    \includegraphics[width=\columnwidth]{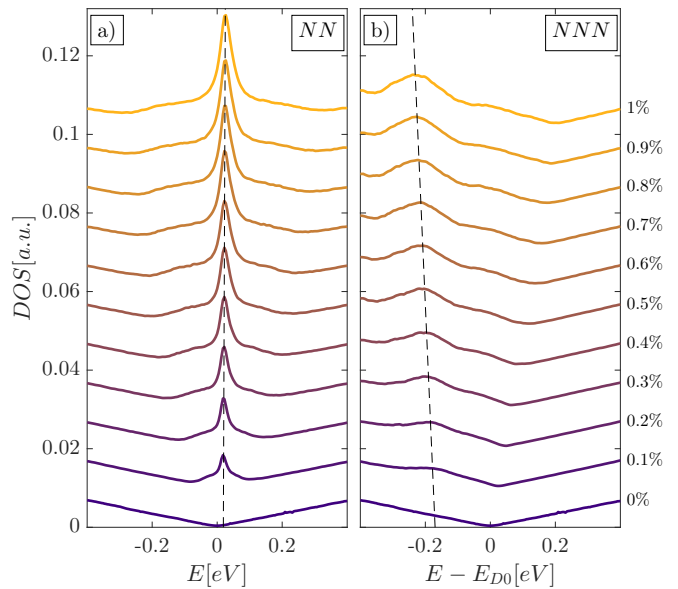}
    	    \caption{\label{figure:dos_vacancies_with_and_without_next_nearest} Low energy DOS for different vacancy concentrations without (NN) a) and with (NNN) b) next-nearest neighbor hopping for $\theta=21.787^\circ$ TBLG. In the latter case states are quasi-localized, in the energy region of the extended states. The vacancy peak when next-nearest neighbors are included has a shift ($\sim 10meV$) which is concentration dependent, while without taking them into account a shift is observed only with respect to the initial Dirac point ($0eV$). Each line is offset with $\Delta y = 0.02$ with respect to the previous one for better distinction.}
        \end{figure}

        Our first conclusion is that the vacancy peak seen in the density of states, shown in Fig.~\ref{figure:dos_vacancies_with_and_without_next_nearest}, becomes broadened and is shifted away from the Dirac point when considering particle asymmetric model Hamiltonian with longer intralayer range hopping. This agrees with previous reports on monolayer graphene with electron-hole asymmetry induced by the inclusion of the next-nearest in-plane neighbor in the TB Hamiltonian\cite{pereira_disorder_2006}. We also report a concentration dependent shift on the order of $\sim 10meV$ of the vacancy peak, see Fig.~\ref{figure:dos_vacancies_with_and_without_next_nearest} for the DOS of $\theta=21.78^\circ$ TBLG, without and with the next-nearest neighbors included in the TB model for different vacancy concentrations, shown respectively in Fig.~\ref{figure:dos_vacancies_with_and_without_next_nearest}(a) and Fig.~\ref{figure:dos_vacancies_with_and_without_next_nearest}(b).
        
\begin{figure}[ttt]
    	    \includegraphics[width=\columnwidth]{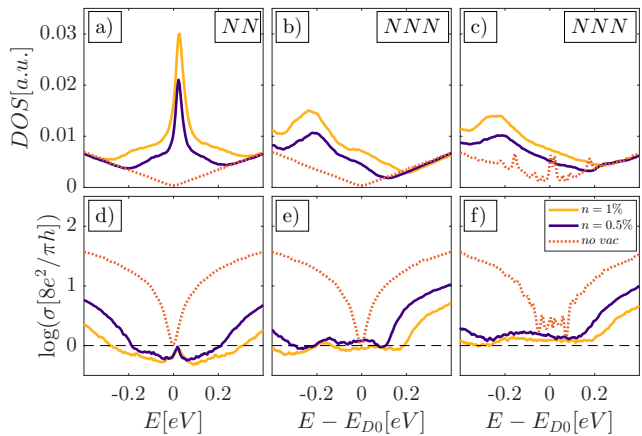}
    	    \caption{\label{figure:Cond_vacancy_TBLG} Comparison of the conductivity and DOS at low energies without vacancies and with vacancy concentration of $0.5\%$ and $1\%$. Results for $\theta = 21.787^\circ$ TBLG are shown in panels  a), b) as  DOS and d), e) as DC conductivity. Results for $\theta = 1.13^\circ$ TBLG are shown in panels c) as DOS and f) as DC conductivity. In panels a) and d) only nearest neighbors are considered (NN), while in other panels the next-nearest neighbors (NNN) are included in the Hamiltonian.}
        \end{figure}

        We investigated the transport properties of TBLG for two different concentrations of vacancy disorder, $0.5\%$ and $1\%$, shown in Fig.~\ref{figure:Cond_vacancy_TBLG}. Similar to the behaviour observed for monolayer graphene with only nearest-neighbor (NN) intralayer hopping, reported in Ref.~[\onlinecite{ferreira_critical_2015}], the vacancies induce localization, concluded from the conductivity calculation. This can be seen in Fig.~\ref{figure:Cond_vacancy_TBLG}(d), where we observe a clear suppression of the conductivity at energies near the Dirac point. The value of conductivity at the neutrality point for $\theta=21.78^\circ$ TBLG is around the value of the sample without vacancies, with a small shift of the Dirac point energy also observed in the DOS, which points to the fact that there are no additional effects due to induced intervalley scattering, and no effects due to the decrease or increase of interlayer tunneling\cite{tao_anti_ferromagnetism_bilayer}. This means that the chiral symmetry is preserved, and that zero energy modes are delocalized. 
        
        We now compare the result with (NNN) and without (NN) the effect of next-nearest neighbor hopping for the same angle. The model Hamiltonian and the nature of the vacancy induced states are very different. Although the DOS shows the resonant states at finite energy, and a different broadening of the vacancy peak, as shown in Fig.~\ref{figure:Cond_vacancy_TBLG}(b), the conductivity has a similar behaviour at the Dirac point. We can distinguish a plateau, shown in Fig.~\ref{figure:Cond_vacancy_TBLG}(e), theoretically described previously in monolayer graphene\cite{katsnelson_vacancy}, which still has the value around the conductivity of the vacancy free sample ($8e^2/\pi h$). The suppression of the conductivity in NNN is weaker around the vacancy peak point, when compared to the short-range (NN) interactions, meaning that these states are not fully localized. Furthermore, based on the values of conductivity, we conclude that at high-angles the physical picture of two decoupled monolayers remains valid even in the presence of strong disorder. 
        
        At low-angles, here $\theta=1.13^\circ$, the universal value of the clean limit of $2\sigma_0$ is not preserved as chiral symmetry is already broken (Fig.~\ref{figure:Cond_rotation}) due to the presence of low energy van Hove singularities or quasi-localized states which can be seen in Fig.~\ref{figure:Cond_vacancy_TBLG}(c), similar to the experimentally confirmed\cite{dombrowski_energy-dependent_2017-1} broken chirality close to the van Hove singularities of monolayer graphene. Vacancies will therefore have a strong effect and suppress the conductivity at all energies, even at the Dirac point, as can be seen Fig.~\ref{figure:Cond_vacancy_TBLG}(f). The low energy van Hove singularities are completely smeared by the presence of the vacancy induced states, and the resulting plateau has similar behaviour and value as in the case of high-twist angles. Upon this, we conclude that distinguishing the transport properties of differently rotated structures of bilayer graphene depends fully on the level of the resonant disorder, which experimentally correspond to the quality of the sample.
\section{\label{section:concl}Conclusions}
        
        In conclusion, we applied a real-space method for the calculation of the DC conductivity to twisted bilayer graphene with a wide range of rotation angles. We examined both high and low twist angles and effects of an applied perpendicular electric field and vacancies. For low rotation angles and in the presence of an electric field, we showed a strong effect of the boundary modes on the DC conductivity, proving that the real-space method clearly captures the change in topology. We see an agreement with previous investigations of the conductivity of vacancy induced ZEM in monolayer graphene and reported results for large angle TBLG. We infer the preservation of chiral symmetry for high-angles, and its breaking in the low twist angle limit. This has important consequences for disordered systems, because the minimal conductivity at the neutrality point is universal only when chiral symmetry is conserved. Our results, showing a nearly flat conductivity due to AB/BA boundary states when a perpendicular electric field is applied is in qualitative agreement with recent experimental findings in low-angle TBLG\cite{kim_tunable_2017}. As the boundary states are expected to survive in the presence of magnetic fields, they will also contribute to the conductivity. These aspects will be presented elsewhere.  
    	    
        \begin{acknowledgments}
            We acknowledge financial support from the graphene FLAG-ERA project TRANS2DTMD.
        \end{acknowledgments}
    	

%

\end{document}